\documentclass[12pt]{article}
\usepackage{longtable,lscape,amssymb}
\usepackage{graphics}
\usepackage{astron2}
\usepackage{afterpage}
\def\degr{\hbox{$^\circ$}}
\def\be{\begin{equation}}
\def\ee{\end{equation}}
\def\bea{\begin{eqnarray}}
\def\eea{\end{eqnarray}}

\def\mkm{\mu{\rm m}}
\def\ga{\mathrel{\mathchoice {\vcenter{\offinterlineskip\halign{\hfil
$\displaystyle##$\hfil\cr>\cr\sim\cr}}}
{\vcenter{\offinterlineskip\halign{\hfil$\textstyle##$\hfil\cr
>\cr\sim\cr}}}
{\vcenter{\offinterlineskip\halign{\hfil$\scriptstyle##$\hfil\cr
>\cr\sim\cr}}}
{\vcenter{\offinterlineskip\halign{\hfil$\scriptscriptstyle##$\hfil\cr
>\cr\sim\cr}}}}}
\def\la{\mathrel{\mathchoice {\vcenter{\offinterlineskip\halign{\hfil
$\displaystyle##$\hfil\cr<\cr\sim\cr}}}
{\vcenter{\offinterlineskip\halign{\hfil$\textstyle##$\hfil\cr
<\cr\sim\cr}}}
{\vcenter{\offinterlineskip\halign{\hfil$\scriptstyle##$\hfil\cr
<\cr\sim\cr}}}
{\vcenter{\offinterlineskip\halign{\hfil$\scriptscriptstyle##$\hfil\cr
<\cr\sim\cr}}}}}
\def\noi{\noindent}
\def\bitem#1#2\par{\noindent\hangindent1.5\parindent\hangafter=1\em#1
\rm#2\par\smallskip}
\def\be{\begin{equation}}
\def\ee{\end{equation}}
\def\bea{\begin{eqnarray}}
\def\eea{\end{eqnarray}}
\begin{document}
\phantom{.}
\bigskip
\bigskip

\begin{center}

{\large\bf SYSTEMATIC VARIATIONS \\
\smallskip
OF INTERSTELLAR LINEAR POLARIZATION AND \\
\smallskip
GROWTH OF DUST GRAINS}
\end{center}

\bigskip

\centerline{\bf {\bf \copyright} 2013 \,\,\,
N.~V.~Voshchinnikov$^{1,\ast}$,
H.~K.~Das$^{2}$,
I.~S.~Yakovlev$^{1}$,
V.~B.~Il'in$^{1,3,4}$
}

\begin{center}
{\it $^1$Sobolev Astronomical Institute
St.~Petersburg University } \\
{\it $^2$ Inter-University Center for Astronomy and Astrophysics,
Pune, India} \\
{\it $^3$Pulkovo Observatory,
St.~Petersburg} \\
{\it $^4$St.~Petersburg State University of Aerospace Instrumentation, 
St.~Petersburg}

\end{center}

\vskip 10pt

\centerline{\small Submitted 21.02.2013}

\begin{quote}\small
\noi
{\bf Abstract.} 
A quantitative interpretation of the observed relation
between the interstellar linear polarization curve parameters
$K$ and $\lambda_{\max}$ characterizing the width and
the wavelength of a polarization maximum, respectively, is given.
 The observational data available for 57 stars located
in the dark clouds in Taurus, Chamaeleon, around the stars $\rho$ Oph and R CrA
are considered.
 The spheroidal particle model of interstellar dust grains
earlier applied to simultaneously interpret the interstellar
extinction and polarization curves in a wide spectral range
is utilized.
 The observed trend $K \approx 1.7 \lambda_{\max}$
is shown to be most likely related to a growth of dust grains due to
coagulation rather than mantle accretion.
 The relation of the parameters $K$ and $\lambda_{\max}$ with
an average size of silicate dust grains is discussed.
\end{quote}


\vfill

\noi \underline{\phantom{...................................}}

\noi{$^{\ast}$E-mail: nvv@astro.spbu.ru}

\newpage

\bigskip
\centerline{ INTRODUCTION}
\medskip

{  The interstellar (IS) polarization is a result of
the linear dichroism of the interstellar medium which is caused
by aligned nonspherical dust grains in the lines of sight. 
 Extinction of the light by such particles depends on
the orientation of the electric vector of the incident radiation. 
 The linear polarization can be described by the 
polarization degree $P$ and the positional angle 
$\theta_{\rm E}$ ($\theta_{\rm G}$) measured in
the equatorial (galactic) coordinate system.
 Historically, the direction of the IS polarization is related to
the direction of the magnetic field component perpendicular
to the line of sight. 

Significant efforts were made to analyze the wavelength dependence of
the IS polarization $P(\lambda)$.
 The polarization degree usually has a maximum in the visual and
monotonically decreases to the ultraviolet and infrared (IR).
 Serkowski~(\cite{serk73}) suggested an empirical formula to describe
the dependence $P(\lambda)$ in the visual 
$$
P(\lambda)/P_{\max} = \exp [-K \ln^2 (\lambda_{\max}/\lambda)].
\label{serkk}
$$
Initially, this equation called Serkowski law
was considered with two parameters:
the maximum polarization wavelength $\lambda_{\max}$
and degree $P_{\max}$,
while the parameter $K$ was taken to be constant 
equal to 1.15 (Serkowski, \cite{serk73}).
 This parameter determines the half-width 
of the normalized IS linear polarization curve
$$W = \lambda_{\rm max}/\lambda_{-} - 
 \lambda_{\rm max}/\lambda_{+},$$
where
$\lambda _{-} < \lambda _{\max } < \lambda _{+}$ and 
$P(\lambda _{+}) = P(\lambda _{-}) = P_{\max }/2$.
The relation between $W$ and $K$ is as follows: 
$$
W = \exp [(\ln 2/K)^{1/2}] - \exp [-(\ln 2/K)^{1/2}].
\label{wk}$$

Using the IR polarization data for 30 stars and 
considering $K$ as a free parameter,
Wilking et al.~(\cite{wilketal82}) found the correlation
$$
K = (1.86 \pm 0.09) \lambda_{\rm max} + (-0.10 \pm 0.05).
$$
Later Whittet et al.~(\cite{wetal92}) reconsidered this correlation
utilizing the observations of 109 stars and obtained
\be
K = (1.66 \pm 0.09) \lambda_{\rm max} + (0.01 \pm 0.05).
\label{k92}
\ee
The linear function (\ref{k92}) well describes the general
dependence of $K$ on $\lambda_{\max}$ derived for
several dark clouds, 
but the observational data for some lines of sight
can essentially differ from this dependence
(Whittet et al., \cite{wetal92}; Voshchinnikov, \cite{v12}).

A qualitative explanation of the relation between the width of
the IS polarization curve and the position of its maximum is connected
to the growth of dust grains in the accretion and coagulation processes
which lead to narrowing of the particle size distribution
(see, e.g., Whittet et al., \cite{wetal92}).
The only attempt to quantitatively interpret the dependence
$K$($\lambda_{\rm max}$) was made by
Aannestad and Greenberg (\cite{ag83}) who considered
effects of the ice mantle growth on cylindrical particles.
 This work has been criticized for
a bad selection of the observational data and an incorrect 
calculation scheme 
(Mathis, \cite{m86}; Voshchinnikov, \cite{v89}; 
Whittet et al., \cite{wetal92}). 

In this work we model the dependence of $K$ on $\lambda_{\rm max}$
for a number of stars located in 4 dark clouds with different
star formation activity.
We use the model of spheroidal particles that was earlier applied 
to simultaneously interpret the IS extinction and polarization
curves in a wide spectral range 
(Voshchinnikov and Das, \cite{vd08}; Das et al., \cite{dvi10}).

\bigskip
\centerline{ OBSERVATIONAL DATA}
\medskip

We have selected the observational data available for 
the dark clouds in Taurus (14 stars), Chamaeleon (22 stars),
around $\rho$ Oph (8 stars), and R CrA (13 stars). 
 All the clouds are  well-known sites of star formation
(Mellinger, \cite{mel08}).
 They are situated in the local interstellar medium at the distance
$D \approx 120 - 140$~pc from the Sun
(Whittet et al., \cite{wetal94}, \cite{wetal01};
Snow et al., \cite{snow08}; Peterson et al., \cite{pet11}). 
 The clouds lie outside the Galaxy plane
$|b| \ga 12\degr$ and either belong to the Gould Belt or
are close to it.

The observational data are collected in Tables 1--4 where
one can find the name of the stars, their galactic coordinates, spectral type,
visual extinction $A_V$ as well as the values of the Serkowski law parameters:
$P_{\max}$, $\lambda_{\max}$, and $K$. 
 The polarization data for the dark cloud in Taurus were 
taken from Efimov (\cite{ef09}) and Whittet et al. (\cite{wetal92}),
for thr dark cloud in Chamaeleon from
Andersson and Potter (\cite{ap07}) and Whittet et al. (\cite{wetal01}),
for the dark cloud around $\rho$ Oph from 
Martin et al. (\cite{maretal92}) and Wilking et al. (\cite{wilketal82}),
and for the dark cloud around R~CrA
from Andersson and Potter (\cite{ap07}) and Whittet et al. (\cite{wetal92}).
 Note that in all the clouds the positional angles are rather ordered. 
 The stars selected are densely situated in the clouds in Chamaeleon
 and around R~CrA and widely distributed (the angular distance between the stars may be of several degrees) in the clouds in Taurus
and around $\rho$ Oph.
 The distances to the stars are mainly about 100--200 pc
and only in a few cases reach 300--500 pc.

\enlargethispage{\baselineskip}

\bigskip
\centerline{ MODELLING}
\medskip

We represent the IS dust grains by homogeneous spheroids
of different size and orientation.
 A solution to the light scattering problem for such particles
has been given by Voshchinnikov and Farafonov (\cite{vf93}).
To compare the theory and observations, one needs to calculate
the intensity of radiation passed through an ensemble of 
partly aligned nonspherical particles.
 Such computations include two steps:
\begin{enumerate}
\item[1)]
calculation of the polarization cross-sections 
$C_{\rm pol}=(C^{\rm TM}_{\rm ext}-C^{\rm TE}_{\rm ext})/2$,
where the superscripts TM and TE denote two orthogonal cases of
orientation of the electric vector of incident radiation 
(Bohren and Huffman, \cite{bh83});

\item[2)]
averaging of these polarization cross-sections over a given
particle distribution over size and orientation. 
\end{enumerate}

Let us unpolarized stellar radiation passes through
a dust cloud with the uniform magnetic field.
 As follows from observations and 
theoretical treatment (Dolginov et al., \cite{dgs79}),
the magnetic field determines the alignment of dust particles.  
 The angle between the line of sight and the magnetic field 
is denoted by $\Omega$ ($0\degr \leq \Omega \leq 90\degr$). 
 The linear polarization produced by rotating spheroidal particles 
in the line of sight is
$$
 P(\lambda)= \sum_j \int\limits_0^D \int\limits_{r_{V,\min,j}}^{r_{V,\max,j}}
    \overline{C}_{{\rm pol},j}(m_{\lambda,j},a_j/b_j,r_{V},\lambda)\,
    n_{j}(r_{V})\, dr_{V} \,dl
    \times 100\,\%\,,
$$
$$
 \overline{C}_{{\rm pol},j}
   = {\frac{2}{\pi^2}}
   {\int\limits_{0}^{\pi/2}}{\int\limits_{0}^{\pi}}{\int\limits_{0}^{\pi/2}}
   \frac{1}{2}
   (C^{\rm TM}_{{\rm ext},j}-C^{\rm TE}_{{\rm ext},j}) \,
    f_j(\xi, \beta, ...) \, \cos 2{\psi} \, d{\varphi}\, d{\omega}\, d{\beta} \,,
\label{cpol}
$$
where $D$ is the distance to the star, 
$\lambda$ the wavelength,
$m_{\lambda,j}$, $a/b_j$ Х $n_{j}(r_{V})$ are 
the refractive index, aspect ratio and size distribution
of spheroidal particles of the $j$th kind, 
$r_{V}$ is the radius of a sphere whose volume is equal
to that of the spheroid 
(for prolate particles, $r_{V} = \sqrt[3]{a b^2}$, and
for oblate ones, $r_{V} = \sqrt[3]{a^2 b}$),
${r_{V,\min,j}}$ and ${r_{V,\max,j}}$ are 
the minimum and maximum radii, 
$C^{\rm TM, \, TE}_{{\rm ext},j}$
the extinction cross-sections for two polarization modes
depending on the particle orientation,
the angle $\psi$ can be expressed through $\varphi, \omega, \beta, \Omega$ 
(see the definitions of the angles and some relations between them,
e.g., in Hong and Greenberg (\cite{hg80}) or Das et al. (\cite{dvi10}),
and finally 
${f}_j(\xi, \beta, ...)$ is the distribution of the particles of the $j$th
kind over orientations.   

We assume that the spheroidal grains are partly aligned so that
their major axes rotate in a plane ($\varphi$ is the angle of rotation)
and their angular momentum {\bf J} precesses around the direction of
the magnetic field ($\omega$ is the precession angle,
$\beta$ the opening angle of the precession cone). 
Such alignment is called the imperfect Davis--Greenstein (IDG) alignment.
 It is described by the distribution function ${f}(\xi, \beta)$
depending only on the orientation parameter $\xi$ and the angle $\beta$.

It should be noted that the problem of dust grain orientation is
the most complicated one in the physics of cosmic dust. 
 In this problem the interactions of dust grains with gas, radiation
 and the magnetic fields are closely related.
 Davis and Greenstein (\cite{dg51}) suggested
that the iron atoms included in dielectric dust grains
made them paramagnetic, which gave the possibility
of the grain interaction with a weak magnetic field.  
 Alignment arises due to the paramagnetic relaxation
of rotating dust grains.
The Davis--Greenstein mechanism was further developed by
Jones and Spitzer (\cite{js67})
who obtained expressions
for the angular momentum distribution function. 
 In the simplest case this function is as follows:
$$
{f}(\xi, \beta) = \frac{\xi \sin \beta}{(\xi^2 \cos^2 \beta + \sin^2 \beta)^{3/2}}.
\label{idg} $$
The parameter $\xi$ depends on the particle size $r_V$,
the imaginary part of the magnetic susceptibility of a dust grain
$\chi''=\varkappa \omega_{\rm d} /T_{\rm d}$,
where $\omega_{\rm d}$ is the angular velocity of the particle,
gas density $n_{\rm g}$, magnetic field strength $B$,
temperatures of dust $T_{\rm d}$ and gas $T_{\rm g}$
$$
\xi^2 = \frac{r_V +\delta_0 (T_{\rm d}/T_{\rm g})}{r_V +\delta_0},
\label{xi} $$
where
$$
\delta_0^{\rm IDG} = 8.23\,10^{23} \frac{\varkappa B^2}{n_{\rm g} T_{\rm g}^{1/2} T_{\rm d}}\,\mkm.
\label{delta} $$
If the particles are not aligned,
we have $\xi=1$ and ${f}(\xi, \beta) = \sin \beta$. 
 In the case of the perfect Davis--Greenstein (PDG) alignment
$\xi=0$ and
${f}(\xi, \beta)=\delta(\beta)$, where $\delta(z)$ is the delta-function.

Different mechanisms of cosmic dust grain alignment 
have been extensively developed in the last few years
(see discussion in Andersson (\cite{and12}) and 
 Voshchinnikov et al. (\cite{vhpd12})),
but their role still remains unclear.

We selected the power-law size distribution of dust grains
}
\be
n(r_V) \propto r_V^{-q}\,
\label{szd}
\ee
introduced by Mathis et al. (\cite{mrn}) from fitting
the IS extinction data.
 This distribution has three parameters:
the minimum ($r_{V, \min}$) and maximum ($r_{V, \max}$) 
{  size}
and the exponent $q$.
 Mathis et al. (\cite{mrn}) have reproduced the mean
IS extinction curve using a mixture of graphite and silicate 
spheres with the parameters:
$q=3.5$, $r_{V, \min}\approx 0.005 \,\mkm$ and $r_{V, \max} \approx 0.25 \,\mkm$.

The ensemble average size of dust grains
$\langle r_{V} \rangle$ is defined as follows:
\be
\langle r_{V} \rangle = \frac{\displaystyle \int\limits_{r_{_{V},\min}}^{r_{_{V},\max}} r_{V} n(r_{V}) \,{d}r_{V}}
{\displaystyle \int\limits_{r_{_{V},\min}}^{r_{_{V},\max}} n(r_{V})\, { d}r_{V}}\,.
\label{rrr}
\ee

In modelling we use the particles of the astronomical silicate (astrosil)
and amorphous carbon (the BE type) whose 
refractive indices were taken from
Draine (\cite{dr03}) and Zubko et al. (\cite{zub96}), respectively.


\bigskip
\centerline{ RESULTS AND DISCUSSION}
\medskip

From Figs. 1--4 (even despite the observational errors)
one can see a clear correlation of the Serkowski law parameters
$K$ and $\lambda_{\max}$ in all the clouds: $K$ increases 
(i.e. the width of the polarization curve $W$ decreases) 
with an increasing $\lambda_{\max}$.
 Slight doubt arises only in the case of the cloud around $\rho$ Oph
where one star strongly affects the correlation coefficient (see Fig.~3).

To interpret the observations we used homogeneous spheroidal particles
of astrosil and amorphous carbon. The prolate and oblate particles with
the size $r_{V} = 0.001 - 0.5\,\mkm$ and the aspect ratio 
$a/b= 1.5, 2, 3, 4, 5$ were utilized.
 Though the silicate and carbonaceous particles comparably
contribute to extinction, the polarization is assumed to be produced only by 
silicate particles.  
Such an assumption has been earlier done by
Chini and Kr\"ugel (\cite{chkru83}) and Mathis (\cite{m86}),
and recently has got additional support 
in the work of Voshchinnikov et al. (\cite{vhpd12})
who found a correlation between the observed IS polarization degree and
the abundance of silicon in dust grains.
\begin{figure}[htb]
\centerline{
\resizebox{12cm}{!}{\includegraphics{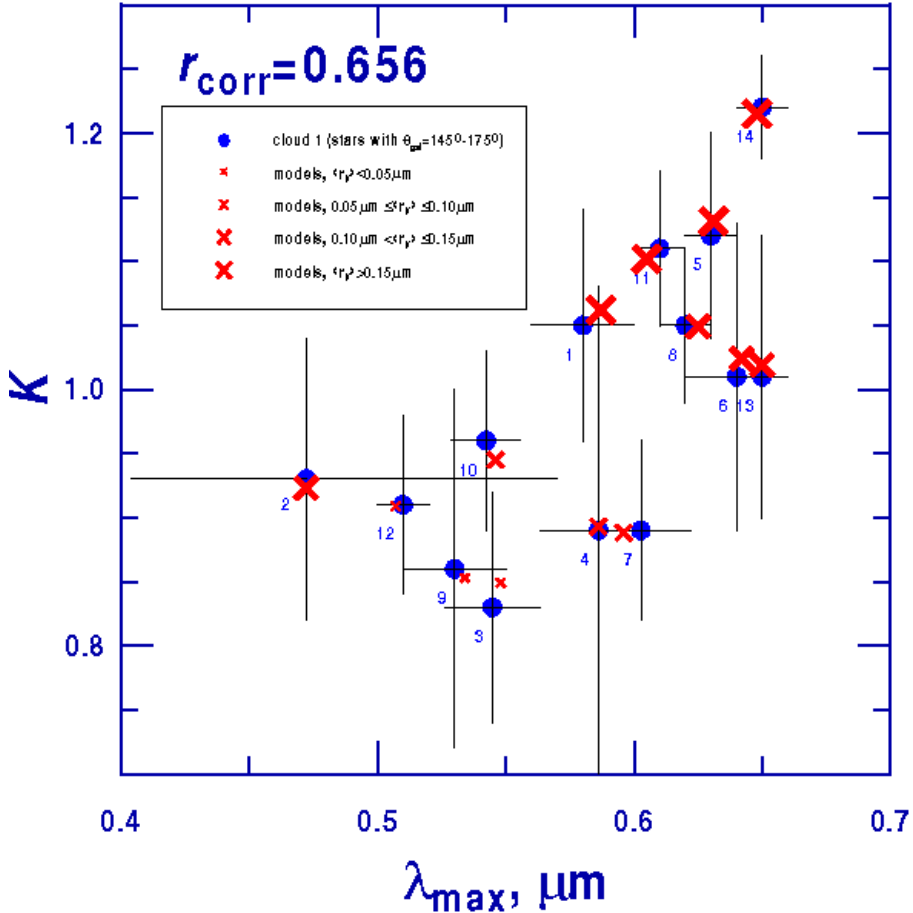}}
}
\caption{ 
The Serkowski law parameter $K$ in dependence on the
maximum polarization wavelength $\lambda_{\max}$. 
 The points show the observational data (with their errors)
for 14 stars in Taurus 
(the star numbers are as in Tabl. 1).
The coefficient of correlation
 between $K$ and $\lambda_{\max}$ is $r_{corr}$.
 Crosses illustrate the results of modelling.
 The size of crosses is proportional to the
mean size of silicate particles $\langle r_{\rm Si} \rangle$.
}
\label{f-t1}
\end{figure}

\begin{figure}[htb]
\centerline{
\resizebox{12cm}{!}{\includegraphics{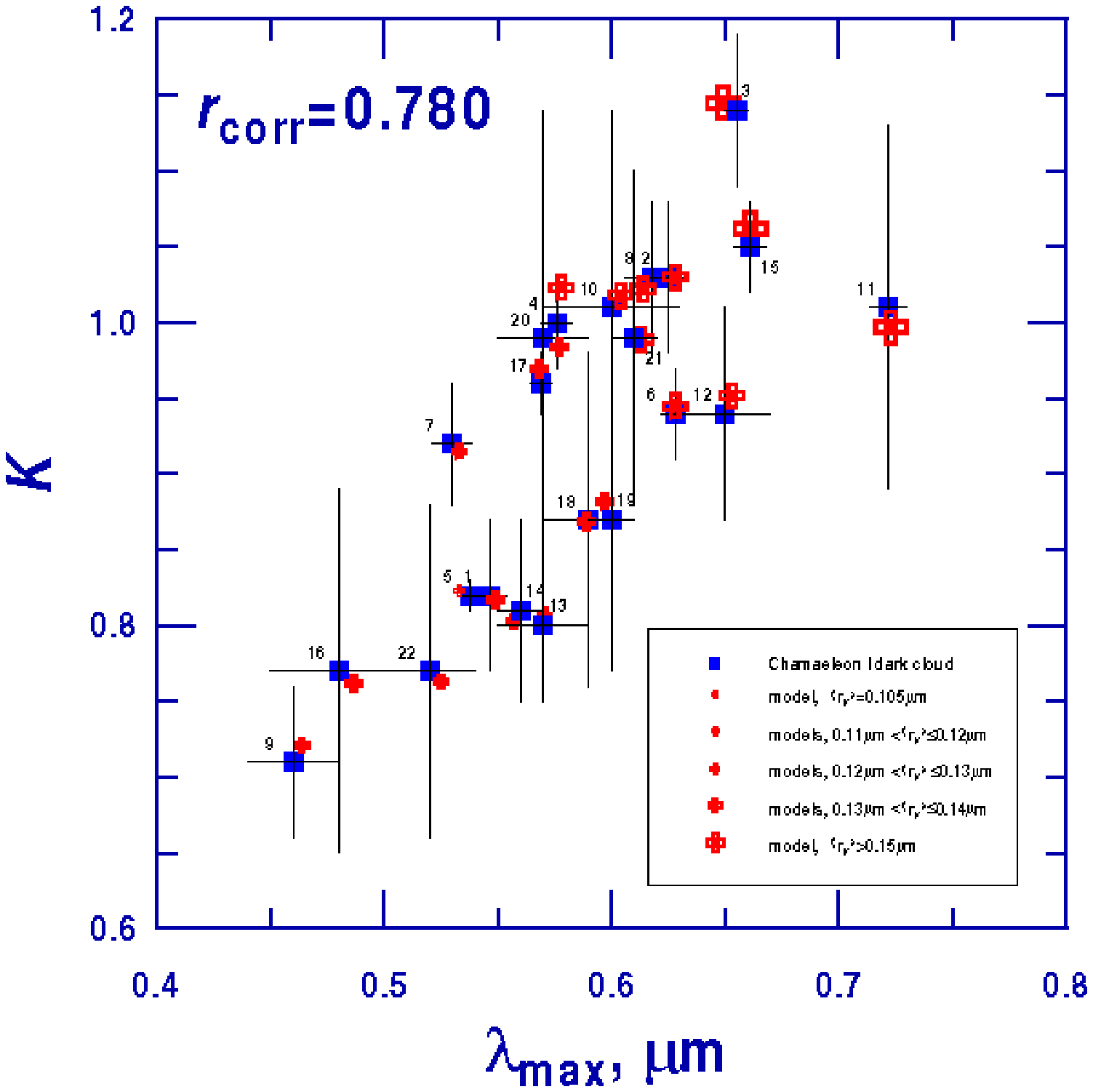}}
}
\caption{ 
The same as in Fig.~\ref{f-t1}, but for 22 stars
in the dark cloud in Chamaeleon (see Tabl.~2). 
}
\label{f-t2}
\end{figure}


For each star in Tabl. 1--4, we have constructed a set of the models,
calculated the Serkowski law parameters: $P_{\max}$, $\lambda_{\max}$,
and $K$, and compared them with the observed value.
 In this paper the discussion is restricted by
consideration of the relation between the width of
the polarization curve and the position of its maximum. 
 These characteristics of the Serkowski law are mainly determined
by the parameters of the size distribution of silicate particles
and weakly depend on the degree and direction of the particle orientation. 
 This orientation is known to mainly affect the ratio $P_{\max}/A_V$
being the polarizing efficiency of the interstellar medium
in a given direction. 
Therefore, 
the ratio $P_{\max}/A_V$ can be used to estimate the spacial structure of
the magnetic fields in interstellar clouds (Voshchinnikov, \cite{v12}).

For all stars, we found the models whose parameters 
$K$ and $\lambda_{\max}$ are close to the observed ones.
These theoretical values of the parameters are represented
by the cross size in Figs. 1--4.
 All the values agree with the observed ones within the errors.
In these models we used spheroids with the aspect ratio $a/b= 3$ or 4,
other ratios gave similar results.
Our search for the models reproducing the observational data
was mainly performed by varying the parameters of the size distribution:
$r_{V, \min}$, $r_{V, \max}$, and $q$.
 In this fitting we kept the total to selective extinction ratio 
$R_V = A_V / E(B-V)$ of our mixtures of silicate and carbonaceous particles
within a reasonable interval. 

\begin{figure}[htb]
\centerline{
\resizebox{12cm}{!}{\includegraphics{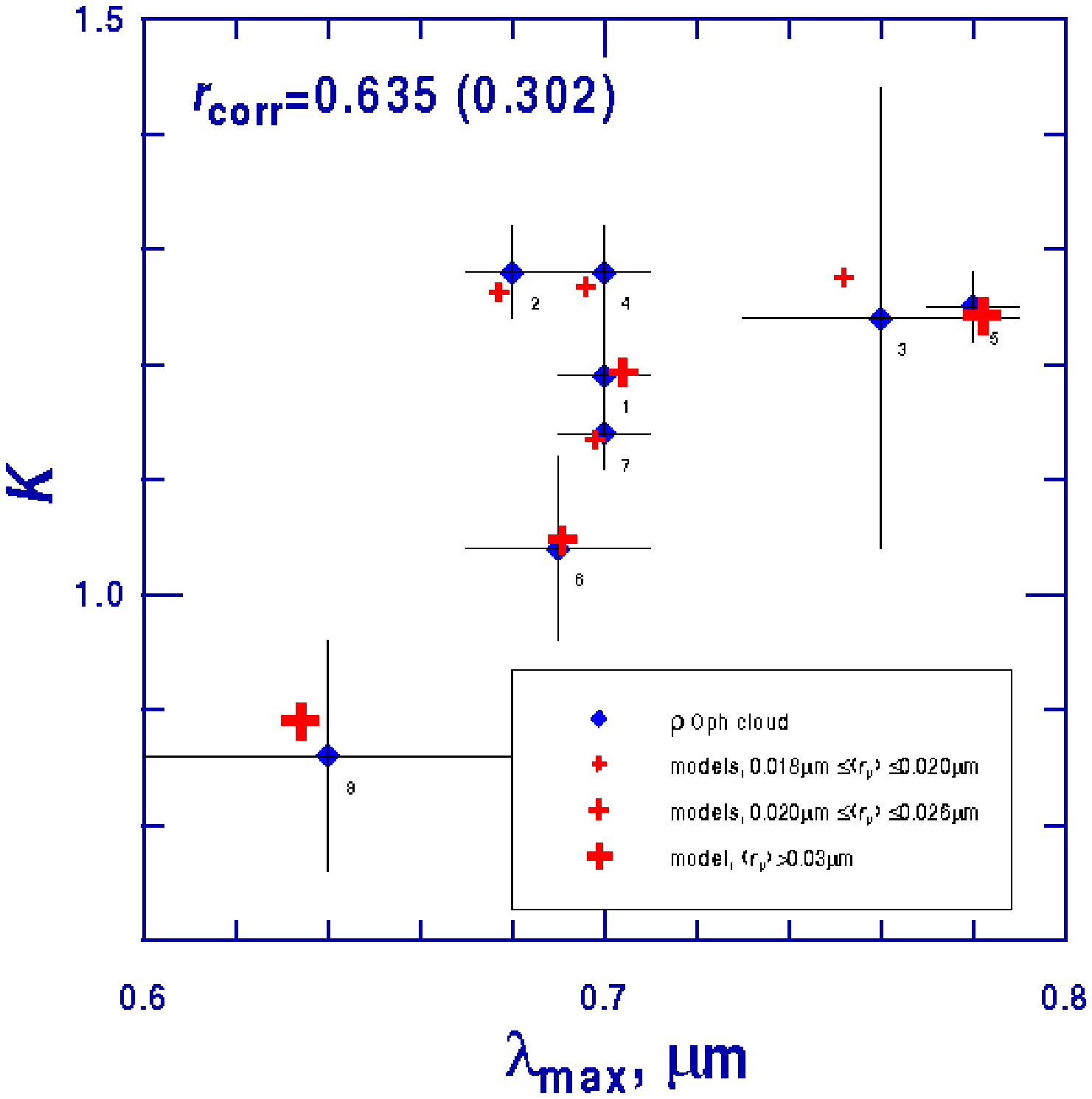}}
}
\caption{ 
The same as in Fig.~\ref{f-t1}, but for 8 stars
in the dark cloud around $\rho$ Oph (see Tabl.~3).  
 The correlation coefficient $r_{corr}$ is computed 
for all stars (in brackets the coefficient for all stars without N~8).
}
\label{f-t3}
\end{figure}

\begin{figure}[htb]
\centerline{
\resizebox{12cm}{!}{\includegraphics{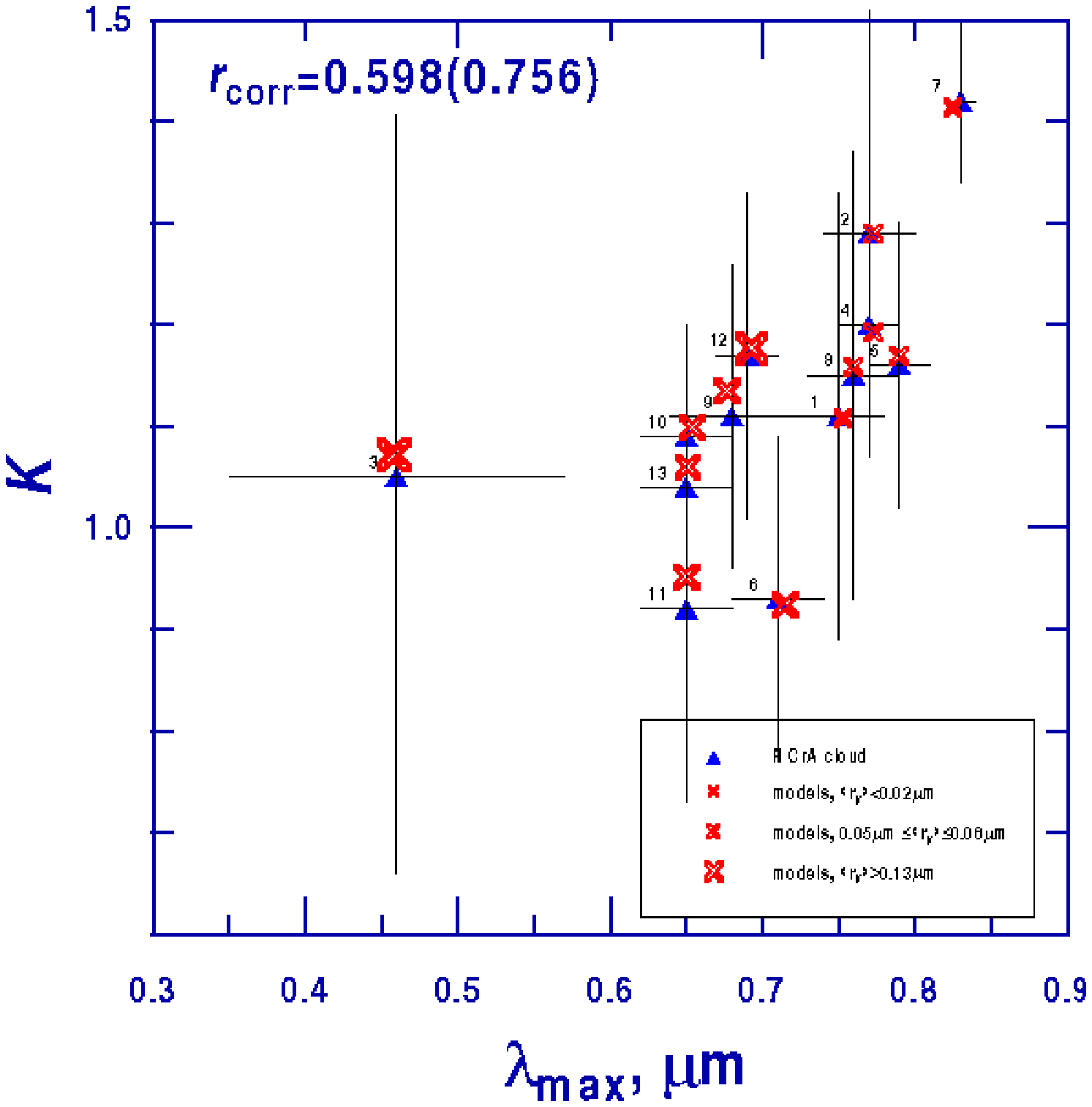}}
}
\caption{  
The same as in Fig.~\ref{f-t1}, but for 13 stars
in the dark cloud around R CrA (see Tabl.~4).  
 The correlation coefficient $r_{corr}$ is computed 
for all stars (in brackets the coefficient for all stars without N~22).
}
\label{f-t4}
\end{figure}

It should be noted that different values of the model parameters
(e.g., those of the particle shape and of the size distribution)
can give similar results ---
this point has been discussed by Das et al. (\cite{dvi10}).
However, in such cases the mean size of particles $\langle r_{V} \rangle$
weakly changes. So, it is a useful parameter 
to characterize the particle ensembles.
 The values of the mean size of silicate particles
 $\langle r_{\rm Si} \rangle$  in the successful models
 are presented in the last column of Tabl. 1--4.
 
Already the first look at the tables 
leads to the conclusion that there is no correlation
between visual extinction $A_V$ and the mean size of silicate particles
$\langle r_{\rm Si} \rangle$.  
The same is correct when we consider the mean size of both
silicate and carbonaceous particles.
However, this cannot be considered as an argument against
the growth of dust grains in the dense parts of the clouds.
 It rather argues for inhomogeneous structure of the clouds and
the presence of extended regions of low density in the lines of sight.

\enlargethispage{\baselineskip}

The mean size of interstellar dust grains 
is thought to be changed mainly by the following processes:  
\begin{enumerate}
\item
{\it Accretion of gas atoms on dust grains.}
 In this case the growth rate is not assumed to 
depend on the particle size  
(Greenberg, \cite{g68}; Voshchinnikov, \cite{v86}).
So, an increase of the minimum and maximum sizes
of dust grains occurs with the exponent being constant,
i.e.  $r_{V, \min}$ grows, $r_{V, \max}$ grows, and $q=$~const. with time.

\item
{\it Destruction of smaller dust grains,}
e.g. because of their evaporation close to hot objects.
 Here we have an increase of the minimum size only
or in other words the size distribution becomes narrower,  
i.e. $r_{V, \min}$ grows, $r_{V, \max}=$~const., and $q=$~const. with time.

\item
{\it Coagulation due to grain-grain collisions.}
 In this case the exponent changes (the size distribution becomes more flat),
i.e. $r_{V, \min}=$~const., $r_{V, \max}=$~const., and $q$ decreases with time.
\end{enumerate}

The scheme described is very simplified\footnote{More detailed consideration
of astrophysical applications of grain growth and destruction in the ISM
can be found in Zhukovska et al. (\cite{zu08}) and Hirashita (\cite{hir12}).},
but it is suitable for understanding of the main effects produced by
the dust grain growth processes mentioned.
 In particular, it is easy to see that
grain coagulation should lead to an anticorrelation
of the parameters $q$ and $\langle r_{\rm Si} \rangle$, while
two other processes should cause
a correlation between $r_{V, \min}$ and $\langle r_{\rm Si} \rangle$.
 These dependencies are presented in Figs. 5 and 6.

From Fig. 5 one can see that 
the grain growth due to coagulation 
may take place in all the clouds but with a different efficiency.
 The growth due to accretion may occur in the cloud in Taurus (see Fig. 6).

Our results show that the mean size of particles producing
the linear polarization in the clouds around $\rho$~Oph and R~CrA
is essentially smaller than that in the clouds in Taurus and Chamaeleon.  
 It can be seen in Figs. 1--4 where the size of crosses presenting
the theoretical results is proportional to the mean size of particles. 
 Note a different position of the regions of smaller and larger 
values of $\langle r_{\rm Si} \rangle$ 
in the figures: small crosses concentrate in the left bottom quarter
for the clouds in Taurus and Chamaeleon and in the right upper
quarter for the clouds around $\rho$~Oph and R~CrA. 
 For the large crosses, the situation is opposite.
 It should be noted that there are stars in the clouds around
$\rho$~Oph and R~CrA  with $\lambda_{\max} > 0.7\,\mkm$, while
for other two clouds usually $\lambda_{\max} \la 0.7\,\mkm$.
 Besides that for some stars with large values of $\lambda_{\max}$
the IS polarization curve is very narrow (the values of $K$ are very
large, see Figs. 3, 4).

\begin{figure}[htb]
\centerline{
\resizebox{12cm}{!}{\includegraphics{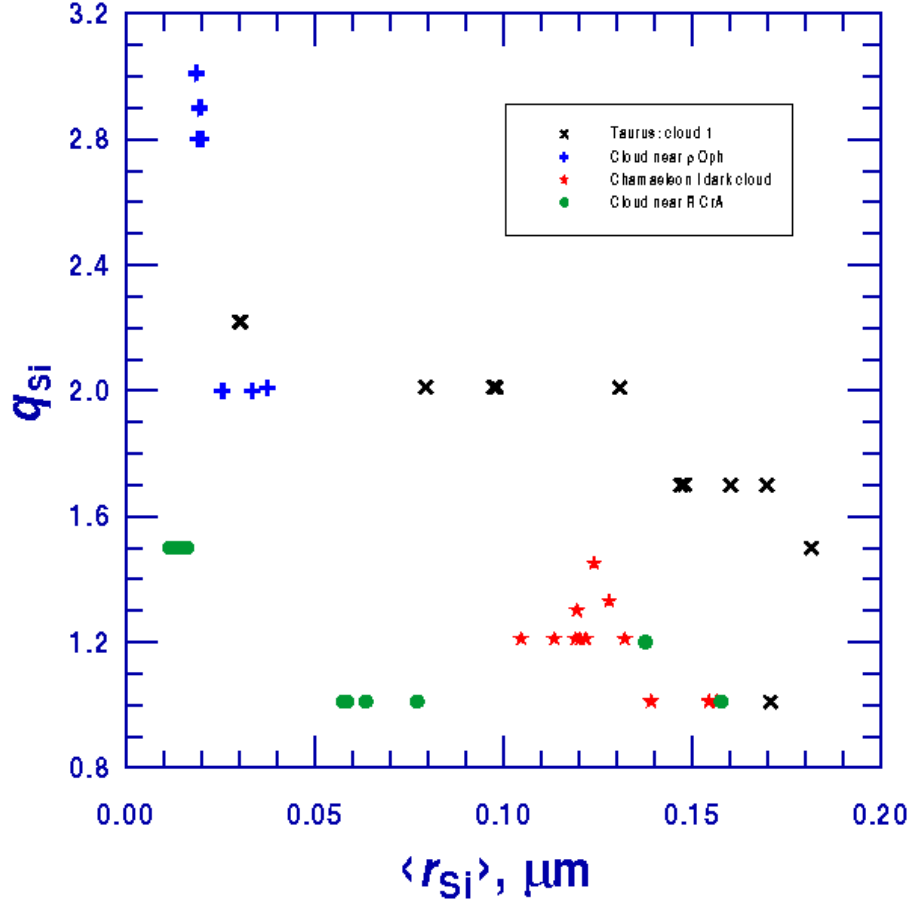}}
}
\caption{  
 The exponent $q_{\rm Si}$ of the silicate particle size 
 distribution  
in dependence on the mean size $\langle r_{\rm Si} \rangle$. 
 The values of $q_{\rm Si}$ and $\langle r_{\rm Si} \rangle$ 
are obtained from fitting of the observational data for
all 57 stars under consideration.
}
\label{f-qr}
\end{figure}

It is interesting to consider the dependence between
the parameter $K$ or $\lambda_{\max}$ found by fitting
and the mean size of particles.
 The dependence of $K$ on $\langle r_{\rm Si} \rangle$
is shown in Fig. 7,
 $\lambda_{\max}$ depends on $\langle r_{\rm Si} \rangle$ in a similar way.
 Both dependencies have non-monotonous character (see Fig. 7).
The polarization curve reaches a maximum width (the minimum value of $K$) 
for $\langle r_{\rm Si} \rangle \approx 0.10\,\mkm$.
 With a growing mean size of particles, the curve $P(\lambda)$
becomes more narrow and $\lambda_{\max}$ increases. 
 However, the extreme values of $\lambda_{\max}$ and $K$
can be also explained using rather small particles.
 Generally, the dependence of $K$ on $\langle r_{\rm Si} \rangle$ or
$\lambda_{\max}$ on $\langle r_{\rm Si} \rangle$ 
can be described by a parabola (see Fig. 7), which
does not correspond to the simple linear function
(see, e.g., Whittet, \cite{w03})
$$
\lambda_{\max} \approx 2 \pi \langle r \rangle (n-1),
$$
where $n$ is the real part of the refractive index.

\begin{figure}[htb]
\centerline{
\resizebox{12cm}{!}{\includegraphics{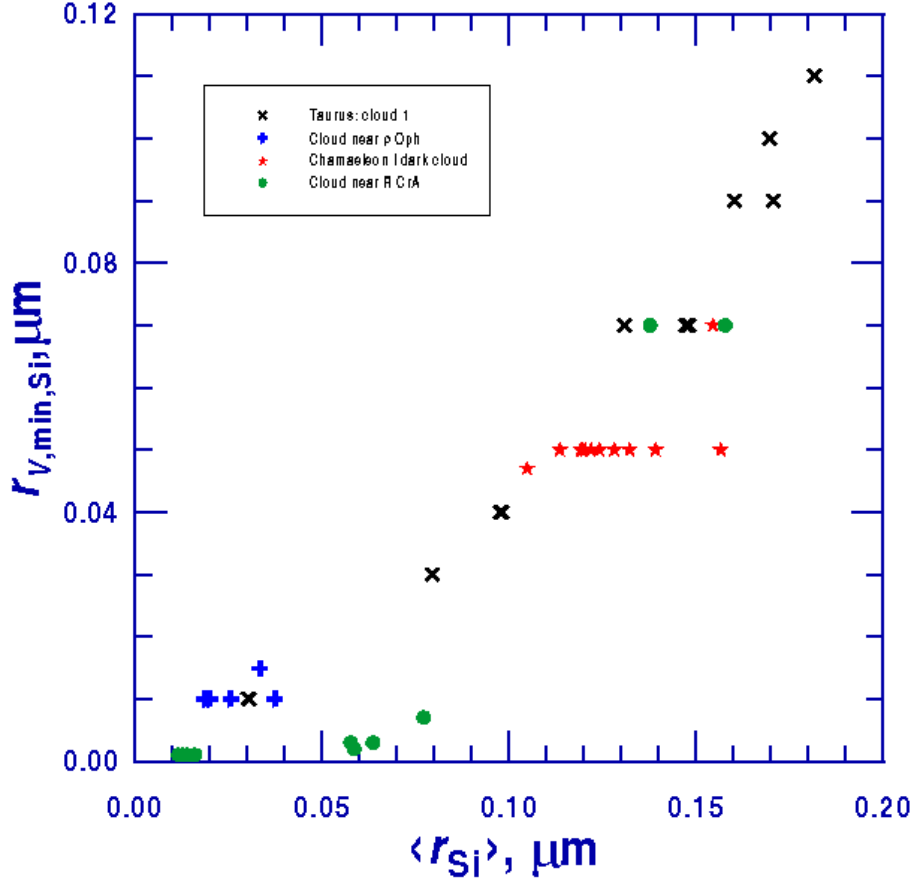}}
}
\caption{The same as in Fig.~\ref{f-qr},
but for the relation between the minimum $r_{V,\min,\rm Si}$
and mean $\langle r_{\rm Si} \rangle$ sizes of silicate particles.
}
\label{f-rminr}
\end{figure}

\begin{figure}[htb]
\centerline{
\resizebox{12cm}{!}{\includegraphics{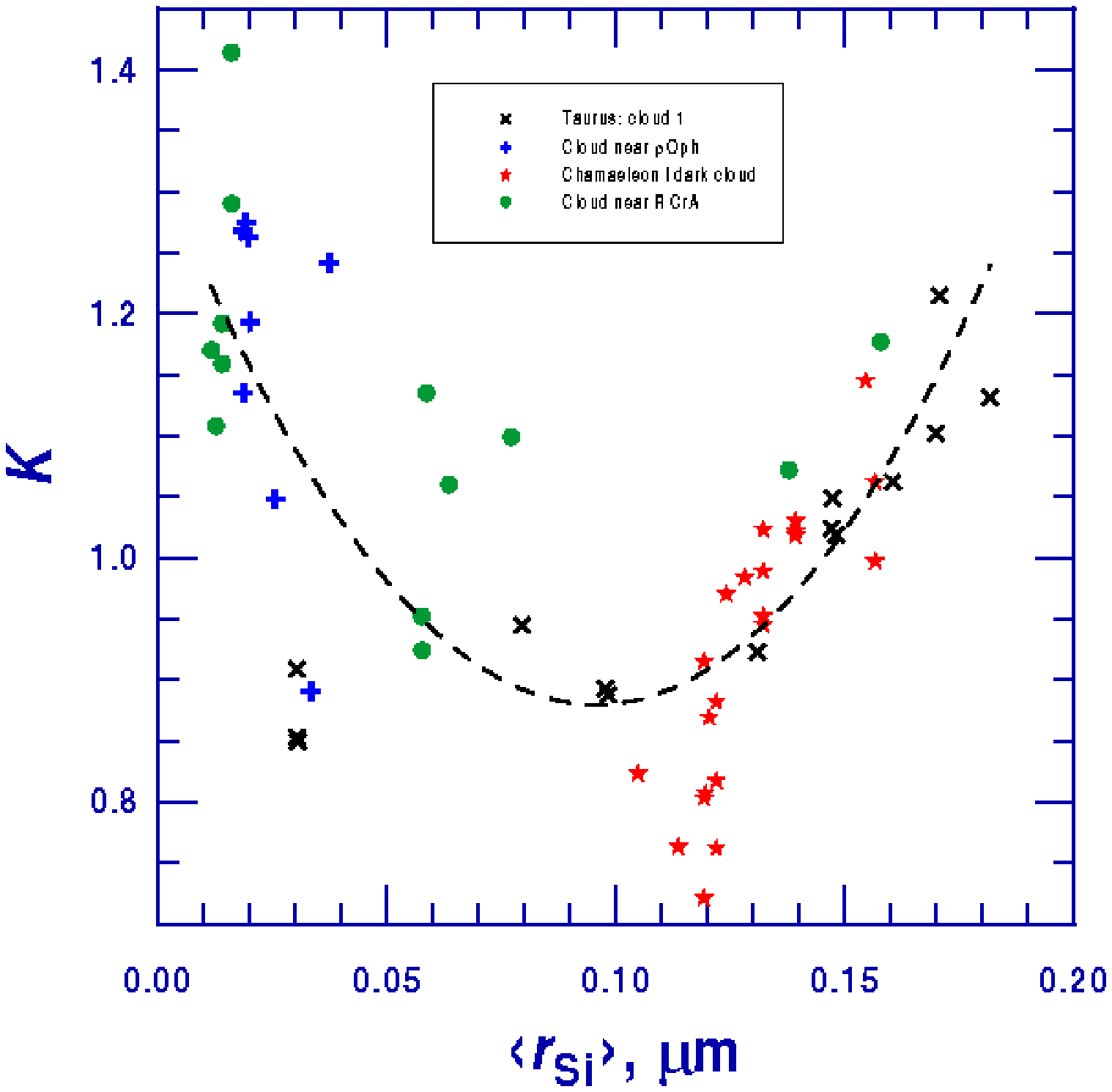}}
}
\caption{ The Serkowski law parameter $K$ 
in dependence on the mean size of silicate particles
$\langle r_{\rm Si} \rangle$.
 The values of $K$ and $\langle r_{\rm Si} \rangle$ were
obtained from fitting the data for 57 stars. 
The dotted line shows the approximation of the dependence of
$K$ on $\langle r_{\rm Si} \rangle$ by a quadric polynomial. 
}
\label{f-kr}
\end{figure}

The polarization data for the clouds around $\rho$~Oph and R~CrA
can be reproduced if one uses the particle ensembles with
$q \la 0$ (see Fig. 8).
In the figure we present the data for all 57 stars under consideration.
 The theoretical dependencies show changes of the Serkowski law
parameters when only accretion 
(for the mechanism 2, the situation is similar) 
or coagulation (the mechanism 3) occurs.  
 The model with the parameters: $a/b = 3$ 
 (prolate spheroids), $r_{V, \min}=0.07\,\mkm$, $r_{V, \max}=0.35\,\mkm$, and
$q=2$ was selected as a basic model.
 As follows from Fig.~8, coagulation of dust grains leads to 
a monotonous growth of $K$ and $\lambda_{\max}$,
but at the same time the wavelength dependence of extinction $A(\lambda)$
becomes less selective and the ratio $R_V$ increases up to very high values.
Considering the accretion process, one should pay attention 
to the non-monotonous character of the dependence
of $K$ on $\lambda_{\max}$,
i.e. similar polarization curves can be derived 
for quite different particle ensembles.
Another interesting feature of the accretion mechanism 
is that for a fixed $q$ one cannot obtain 
the values of $\lambda_{\max}$ below some limit value
(e.g., for $q=2$ we got $\lambda_{\max} \ga  0.51\,\mkm$).
 Obviously, the possibility of using smaller particles to interpret
the data for stars with very large values of $\lambda_{\max}$
needs further analysis and discussion.   
 Another way to solve the problem of very narrow $P(\lambda)$ curves
is to apply a mixture of particles of quite different shapes.

\begin{figure}[htb]
\centerline{
\resizebox{12cm}{!}{\includegraphics{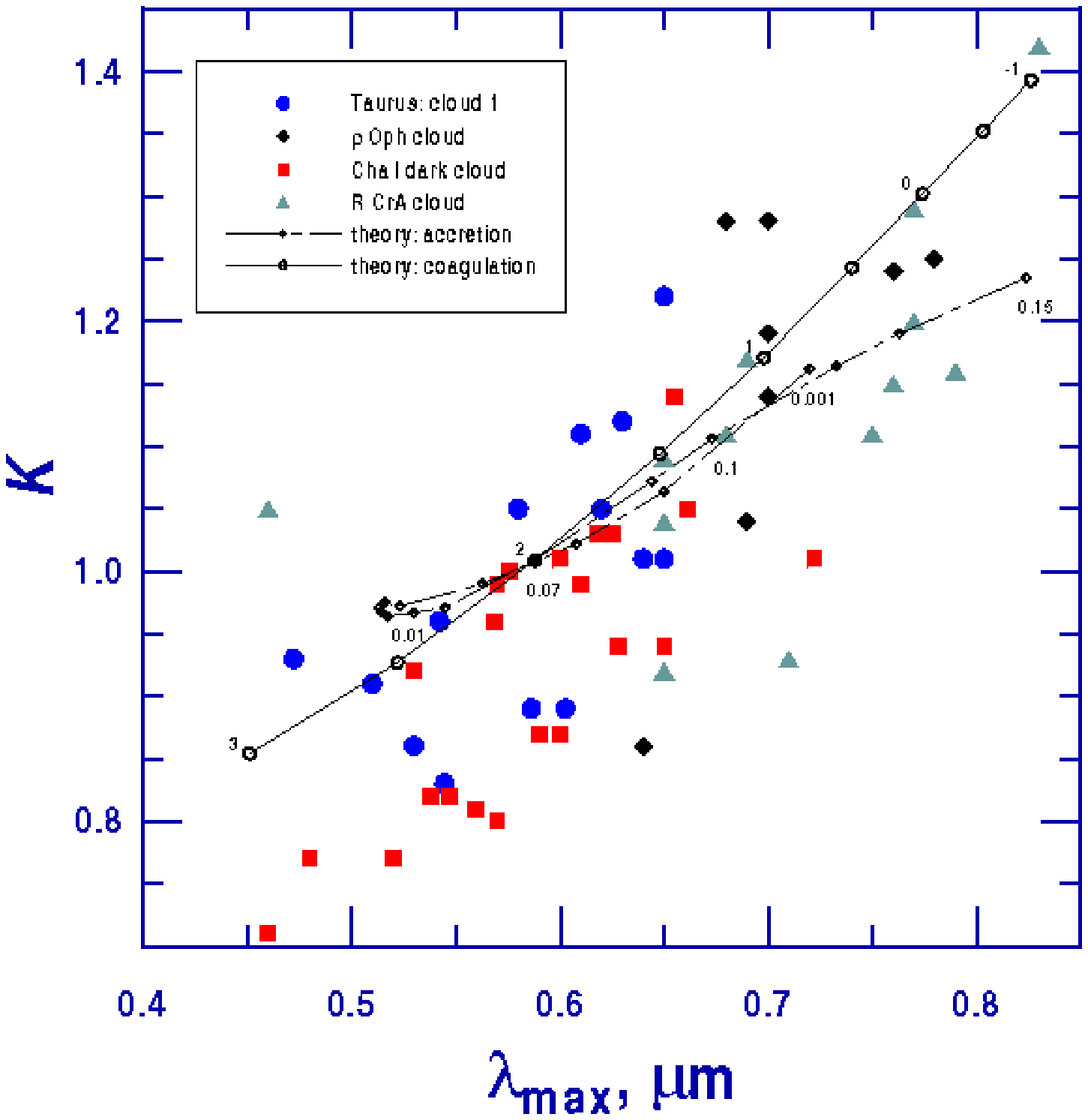}}
}
\caption{ 
The parameter $K$  
in dependence on
$\lambda_{\max}$.
 The filled signs show the observational data (without observational errors)
for 57 stars.
 The empty signs correspond to results of calculations
when the grain growth due to accretion (squares) or coagulation (circles)
takes place.
 The numbers close to these signs show the values of the
exponent $q_{\rm Si}$ or the minimum size $r_{V,\min,\rm Si}$.
The calculations were performed for the angle between
the line of sight and the magnetic field direction 
$\Omega=60\degr$.
}
\label{f-kl}
\end{figure}


\bigskip
\centerline{ CONCLUSIONS}
\medskip

{For 57 stars in 4 dark clouds with different
star formation activity, we have modeled
the observed wavelength dependencies of
the linear polarization degree $P(\lambda)$ described
by the Serkowski law parameters $P_{\max}, \lambda_{\max}$ and $K$. 
 Polarization in the model was produced by
an ensemble of homogeneous silicate spheroidal particles
with a power-law size distribution and imperfect
Davis--Greenstein alignment.  

For all the stars we found the size distribution parameters
giving the observed values of $P_{\max}, \lambda_{\max}$, and $K$.
Analysis of the results obtained has shown that
the dust grain growth can occur in all the clouds under consideration,
but with different efficiency. The growth caused by accretion
may take place only in the cloud in Taurus.

\enlargethispage{\baselineskip}

It is shown that in most of the cases the dust grain ensembles can be
characterized by the only parameter --- the mean size of particles.
 We have considered the relation between the parameters $\lambda_{\max}$, $K$
and the mean size of silicate particles $\langle r_{\rm Si} \rangle$.
It is found that narrow red-shifted polarization curves $P(\lambda)$
observed for some stars can be explained by larger
or smaller particles, i.e. by using ensembles 
with a larger or smaller value of $\langle r_{\rm Si} \rangle$.
These results require a special discussion with regard to the properties
of individual clouds that will be done in a separate paper.

\smallskip

The work was partly supported by the grant of Russian Foundation for Basic Researches
11-02-92695-IND-a.

\afterpage{\clearpage}

\newpage
\begin{table}
\noi {\bf  Table 1.} {Stars in the cloud 1 in Taurus}
\begin{center}
{\small
\begin{tabular}{cccccccccccc}
\noalign{\smallskip}
\hline
\noalign{\smallskip}
$N$ & Star & $l$ & $b$ & Sp.Type & $A_V$ &
$P_{\max},\,\%$ & $\lambda_{\max}$,\,{$\mkm$}& $K$ &
$\langle r_{\rm Si} \rangle $,\,{$\mkm$} 
  \\
\noalign{\smallskip}
\hline
\noalign{\smallskip}
~1 & HD 28225 & 170.74 & --14.35 &A3 III & 1.22 & 1.88 & 0.58 & 1.05 & 0.1604 \\
~2 & HD 29835 & 174.14 & --12.86 &K2 III & 1.20 & 4.06 & 0.472 & 0.93 & 0.1309 \\
~3 & HD 30168 & 174.76 & --12.44 &B8 V & 1.02 & 4.08 & 0.545 & 0.83 & 0.0305 \\
~4 & HD 283637 & 170.51 & --14.84 &A0 V & 2.28 & 2.73 & 0.586 & 0.89 & 0.0976 \\
~5 & HD 283642 & 171.55 & --15.35 &A3 V & 2.19 & 2.01 & 0.63 & 1.12 & 0.1817 \\
~6 & HD 283643 & 171.83 & --15.07 &A2 V & 1.66 & 1.35 & 0.64 & 1.01 & 0.1471 \\
~7 & HD 283701 & 172.18 & --13.63 &B8 III & 2.53 & 3.20 & 0.603 & 0.89 & 0.0982 \\
~8 & HD 283757 & 174.01 & --14.81 &A5 V & 1.65 & 2.92 & 0.62 & 1.05 & 0.1471 \\
~9 & HD 283800 & 173.57 & --12.29 &B5 V & 1.64 & 3.95 & 0.53 & 0.86 & 0.0303 \\
10 & HD 283812 & 174.90 & --13.07 &A1 V & 1.92 & 6.26 & 0.542 & 0.96 & 0.0795 \\
11 & HD 283815 & 175.32 & --13.90 &A5 V & 1.91 & 2.86 & 0.61 & 1.11 & 0.1698 \\
12 & HD 283855 & 174.25 & --11.47 &A2 & 1.99 & 5.13 & 0.51 & 0.91 & 0.0303 \\
13 & HD 283877 & 174.94 & --12.72 &F5 V & 0.72 & 1.65 & 0.65 & 1.01 & 0.1481 \\
14 & HD 283879 & 175.72 & --12.61 &B5 V & 3.33 & 4.24 & 0.65 & 1.22 & 0.1707 \\
\noalign{ \smallskip}
\hline
\end{tabular}}
\end{center}
\end{table}

\newpage
\noi {\bf  Table 2.} {Stars in the dark cloud Chamaeleon I}
\begin{center}
{\small
\begin{tabular}{c|c|c|c|c|c|c|c|c|c}
\noalign{\smallskip}
\hline
\noalign{\smallskip}
$N$ & Star & $l$ & $b$ & Sp.Type & $A_V$ &
$P_{\max},\,\%$ & $\lambda_{\max}$,\,{$\mkm$}& $K$ &
$\langle r_{\rm Si} \rangle $,\,{$\mkm$} 
\\
\noalign{\smallskip}
\hline
\noalign{\smallskip}
~1 & Cha F1 & 296.05 & --15.70 & K4 III & 0.8 & 3.35  & 0.547 & 0.82  & 0.1219 \\
~2 & Cha F2 & 296.65 & --16.60 & B8 V & 1.8 & 3.85 & 0.625  & 1.03  & 0.1392 \\
~3 & Cha F3 & 296.26 & --15.84 & B4 V & 2.4 & 5.45  & 0.655 & 1.14  &0.1545 \\
~4 & Cha F6 & 296.39 & --15.48 & A2 V & 1.6 & 5.48  & 0.576  & 1.00  &0.1322 \\
~5 & Cha F7 & 296.28 & --15.24 & B5 V & 1.7 & 5.92  & 0.538  & 0.82  &0.1048 \\
~6 & Cha F9 & 296.66 & --15.62 & K0 III & 2.5 & 4.82  & 0.628  & 0.94  &0.1322 \\
~7 & Cha F11 & 296.53 & --15.06 & B9 V & 2.9 & 4.81  & 0.530  & 0.92  &0.1192 \\
~8 & Cha F16 & 296.99 & --15.74 & G2 IV & 3.1 & 7.30 & 0.618  & 1.03  &0.1392 \\
~9 & Cha F21 & 296.83 & --15.29 & K3 III & 2.2 & 5.41  & 0.46  & 0.71  &0.1192 \\
10 & Cha F25 & 297.13 & --15.54 & G8 III & 5.4 & 8.01 & 0.60 & 1.01  &0.1392 \\
11 & Cha F28 & 297.24 & --15.73 & K4 III & 6.1 & 7.02 & 0.722 & 1.01  &0.1566 \\
12 & Cha F29 & 297.28 & --15.74 & K6 & 2.7 & 5.05 & 0.65 & 0.94 &0.1322 \\
13 & Cha F30 & 297.04 & --15.13 & K3 III & 1.9 & 4.41 & 0.57 & 0.80  &0.1195 \\
14 & Cha F32 & 296.98 & --14.80 & A7 V & 2.2 & 2.34 & 0.56 & 0.81  &0.1192 \\
15 & Cha F36 & 297.47 & --15.73 & K0 III & 5.7 & 12.19 & 0.661 & 1.05  &0.1566 \\
16 & Cha F39 & 297.49 & --15.20 & K3 III & 0.9 & 3.22 & 0.48 & 0.77  &0.1219 \\
17 & Cha F40 & 297.69 & --15.64 & B8 III & 2.1 & 8.01 & 0.569 & 0.96  &0.1241 \\
18 & Cha F41 & 296.88 & --13.48 & B8 V & 0.9 & 2.60 & 0.59 & 0.87  &0.1203 \\
19 & Cha F42 & 297.08 & --13.61 & A3/A4 IV& 1.2 & 2.87 & 0.60 & 0.87  &0.1219 \\
20 & Cha F48 & 297.56 & --14.07 & B9.5 V & 0.6 & 2.29 & 0.57 & 0.99  &0.1281 \\
21 & Cha F52 & 297.77 & --13.91 & B9.5 V & 1.2 & 2.99 & 0.61 & 0.99  &0.1322 \\
22 & Cha F54 & 298.17 & --14.17 & G6 III/IV& 0.8 & 2.68 & 0.52 & 0.77  &0.1136 \\
\noalign{ \smallskip}
\hline
\end{tabular}}
\end{center}


\newpage
\noi {\bf  Table 3.} {Stars in the dark cloud around $\rho$ Oph}
\begin{center}
{\small
\begin{tabular}{c|c|c|c|c|c|c|c|c|c}
\noalign{\smallskip}
\hline
\noalign{\smallskip}
$N$ & Star & $l$ & $b$ & Sp.Type & $A_V$ &
$P_{\max},\,\%$ & $\lambda_{\max}$,\,{$\mkm$}& $K$ &
$\langle r_{\rm Si} \rangle $,\,{$\mkm$} \\
\noalign{\smallskip}
\hline
\noalign{\smallskip}
1 & HD 145502 & 354.61 & +22.70 & B2IV & 1.06 & 1.25 & 0.70 & 1.19 & 0.0201 \\
2 & HD 147084 & 352.33 & +18.05 & A4II/III & 2.70 & 4.42 & 0.68 & 1.28 & 0.0197 \\
3 & HD 147283 & 352.29 & +17.61 & A1IV & 2.53 & 1.61 & 0.76 & 1.24 & 0.0192 \\
4 & HD 147888 & 353.65 & +17.71 & B3V:SB & 2.08 & 3.63 & 0.70 & 1.28 & 0.0187 \\
5 & HD 147889 & 352.87 & +17.04 & B2V & 4.44 & 4.02 & 0.78 & 1.25 & 0.0375 \\
6 & HD 147932 & 353.72 & +17.71 & B5V & 2.10 & 3.11 & 0.69 & 1.04 & 0.0256 \\
7 & HD 147933 & 353.68 & +17.70 & B1.5V & 2.07 & 2.69 & 0.70 & 1.14 & 0.0187 \\
8 & HD 150193 & 355.60 & +14.83 & A1Ve & 1.79 & 5.10 & 0.64 & 0.86 & 0.0335 \\
\noalign{ \smallskip}
\hline
\end{tabular}}
\end{center}


\bigskip
\bigskip
\noi {\bf Table 4.} {Stars in the dark cloud around R CrA}
\begin{center}
{\small
\begin{tabular}{c|c|c|c|c|c|c|c|c|c}
\noalign{\smallskip}
\hline
\noalign{\smallskip}
$N$ & Star & $l$ & $b$ & Sp.Type & $A_V$ &
$P_{\max},\,\%$ & $\lambda_{\max}$,\,{$\mkm$}& $K$ &
$\langle r_{\rm Si} \rangle $,\,{$\mkm$} \\
\noalign{\smallskip}
\hline
\noalign{\smallskip}
~1 & RCrA 12 & 359.46 & --18.65 & G8III & 1.7 & 0.81 & 0.75 & 1.11 & 0.0126 \\
~2 & RCrA 15 & 359.65 & --18.66 & G1 & 1.1 & 3.00 & 0.77 & 1.29 & 0.0161 \\
~3 & RCrA 22 & 000.10 & --18.09 & K5III & 3.3 & 1.30 & 0.46 & 1.05 & 0.1377 \\
~4 & RCrA 28 & 000.12 & --17.59 & M5III & 1.3 & 2.10 & 0.77 & 1.20 & 0.0141 \\
~5 & RCrA 30 & 359.52 & --18.04 & A0V & 2.6 & 1.85 & 0.79 & 1.16 & 0.0116 \\
~6 & RCrA 43 & 359.49 & --17.25 & K0III & 1.9 & 1.73 & 0.71 & 0.93 & 0.0577 \\
~7 & RCrA 46 & 359.65 & --17.56 & G8III & 3.3 & 2.71 & 0.83 & 1.42 & 0.0161 \\
~8 & RCrA 50 & 359.44 & --17.37 & A6V & 1.3 & 1.11 & 0.76 & 1.15 & 0.0141 \\
~9 & RCrA 52 & 359.44 & --17.32 & G5III & 1.6 & 1.99 & 0.68 & 1.11 & 0.0586 \\
10 & RCrA 56 & 000.29 & --18.68 & G5IV & 1.9 & 2.08 & 0.65 & 1.09 & 0.0772 \\
11 & RCrA 58 & 359.41 & --17.90 & K1III & 2.3 & 0.80 & 0.65 & 0.92 & 0.0577 \\
12 & RCrA 71 & 000.11 & --18.84 & F6V & 1.3 & 1.03 & 0.69 & 1.17 & 0.1578 \\
13 & RCrA 73 & 000.10 & --19.04 & G0V & 1.2 & 1.80 & 0.65 & 1.04 & 0.0636 \\
\noalign{ \smallskip}
\hline
\end{tabular}}
\end{center}

\end{document}